\documentclass[reprint,amsmath,amssymb,aps,]{revtex4-1}

\usepackage{graphicx}
\usepackage{dcolumn}
\usepackage{bm}
\usepackage{hyperref}
\usepackage[mathlines]{lineno}
\usepackage{physics}

\begin{document}

\title{\bf Wigner distribution and Boltzmann-Shannon entropy  for a diatomic molecule under polarized electric field interaction }
\author{Gustavo V. L\'opez}
 \altaffiliation[]{gulopez@cencar.udg.mx}
\author{Alejandro P. Mercado}%
 \email{pamp.fis@gmail.com}
\affiliation{%
 Departamento de F\'{i}sica, Universidad de Guadalajara,\\
 Blvd. Marcelino Garc\'{i}a Barragan y Calzada Ol\'{i}mpica, \\44200 Guadalajara, Jalisco, M\'exico}

\date{\today}

\begin{abstract}
\vskip1pc\noindent
We study the classical chaos appearing in  a diatomic molecules $BeO$, $CO$ and $CN$ due to the interaction with a circularly polarized electric field, and its signature in Quantum Mechanics through the Wigner distribution function and the Boltzmann-Shannon entropy. We found a motion out of the center of the quantum phase space defined by Wigner function when the classical system becomes chaotic, and we found a jumping behavior of the average  Boltzmann-Shannon entropy with respect the electric field strength when the classical system becomes chaotic, indicating a sudden increasing in the disorder (or sudden lost of information) in the quantum system.       
Ê
\end{abstract}

\pacs{Valid PACS appear here}
\maketitle


\section{\label{sec:level1}Introduction}

The study of dynamical chaos in atomic an molecular systems has been of great theoretical and experimental interest  \cite{1,2,3,4,5,6,7,8,9,10} since not enough integrals of motion are found either in classical or quantum systems. Most of the classical \cite{11,12,13} and quantum \cite{14,15,16} approaches use the Morse potential as a model of interatomic interaction \cite{17}, and a summary of the classical properties of diatomic molecules can be found on reference \cite{17}. The chaotic behavior of diatomic molecules due to double non linear resonances in the action-angle variable produced by the interaction with a circularly polarized electric field has been studied \cite{18,19}. Other approaches to the study of chaotic behavior on diatomic molecules have been used \cite{20,21,22,23,24}  to determine different  aspects of the system. The most looked signature in quantum mechanics for a chaotic behavior in the classical system has been the distribution of first neighbor energies (or quasi-energies) of the quantum system (GOE, GUE, or GSE, depending on the symmetry of the Hamiltonian, for non integrable chaotic case, and Poisson distribution for integrable case) and the so called $\Delta_3$ rigidity \cite{25,26,27}. However, one needs to mention that the quantum harmonic oscillator and quantum rotator are integrable quantum systems without Poisson distribution function on their eigenvalues.  
\\ \\
In this paper, we want to study firstly the classical behavior of a diatomic molecule under a circularly polarized electric field to determine the magnitude of the intensity for the system to appear a chaotic behavior. This will be done with the exact Morse's potential and its approximation at fourth order. Secondly, using this Morse's potential approximation a forth order, we make the quantization of the system through  Schr\"odinger's equation, writing the wave function in terms of the basis of the harmonic oscillator and angular momentum functions, and solve numerically the resulting complex dynamical system for the ground state and several exited states of the system. Finally, we calculate the Wigner function \cite{28} and Boltzmann\cite{29}-Shannon\cite{30} (BS) entropy with this solution, looking their behavior as a function of the amplitude of the electric field strength to see a quantum signature of the classical chaotic system. 

\section{Classical Analysis}
The Hamiltonian of a diatomic molecule interacting with a circular polarized electric field is given by [15]

\begin{equation}\label{ha1}
H=\frac{P_r^2}{2\mu}+\frac{1}{2\mu r^2}\left(P_{\theta}+\frac{P_{\varphi}^2}{\sin^2\theta}\right)+U(r)-{\bf p}\cdot{\bf E},
\end{equation}

\noindent
where $\mu$ is the reduced mass of the molecule, ($r,\theta,\varphi$) are the spherical coordinates of the relative position of the two atoms ($r=|{\bf r}|$), ($P_r,P_{\theta}, P_{\varphi}$) represent their linearly generalized momenta, ${\bf p}=q{\bf r}$ is the dipole moment of the molecule ($q$ is the charge), 
${\bf E}=(E_0\cos\omega t, E_0\sin\omega t, 0)$ is the circularly polarized electric field with frequency $\omega$ and amplitude $E_0$, and $U(r)$ is the Morse's potential,

\begin{equation}\label{Morse}
U(r)=D\bigl[1-e^{-a(r-r_0)}\bigr]^2,
\end{equation}
\noindent
with $D$, $a$, and $r_0$ being constants which depend on the molecule. The radius $r_0$ is such that $U(r_0)=U'(r_0)=0$, $U''(r_0)=2a^2D$, $U'''(r_0)=-6a^3D$, and $U^{iv}(r_0)=14a^4D$. Using the new variable $\xi=r-r_0$ and since $P_{\theta}^2+P_{\varphi}^2/\sin^2\theta=L^2$ represents the total angular momentum of the molecule, the Hamiltonian (\ref{ha1})  and its approximation at fourth order can be written as

\begin{eqnarray}\label{ha2}
H=&\frac{P_{\xi}^2}{2\mu}+\frac{L^2}{2\mu(\xi+r_0)^2}+D\bigl[1-e^{-a\xi}\bigr]^2\\
  &-q(r_0+\xi)E_0\sin\theta\cos(\varphi-\omega t)\nonumber
\end{eqnarray}

\noindent
and

\begin{eqnarray}\label{ha4}
& &H_4=\frac{P_{\xi}^2}{2\mu}+\frac{1}{2}\mu\omega_o^2\xi^2+
\frac{L^2}{2\mu r_0^2}\biggl[1-\frac{2\xi}{r_0}+\frac{3\xi^2}{r_0^2}-\frac{4\xi^3}{r_0^3}+\frac{5\xi^4}{r_0^4}\biggr]\nonumber\\ 
& &\quad -a^3D\xi^3+\frac{7}{12}a^4D\xi^4-W\xi\sin\theta\cos(\varphi-\omega t),
\end{eqnarray}

\noindent
where $W=qE_0$ is the electric field strength, and the time average term \hfil\break $-E_0\langle r_0q\sin\theta\cos(\varphi-\omega t)\rangle$ has been absorbed in the definition of this Hamiltonian [18]. The Hamiltonian dynamical systems generated by (\ref{ha2}) and (\ref{ha4}) are

\begin{subequations}
\begin{eqnarray}         
& &\dot\xi=P_{\xi}/\mu \\ 
& &\dot P_{\xi}=\frac{L^2}{\mu(\xi+r_0)^3}-2aD\bigl[1-e^{-a\xi}\bigr]e^{-a\xi}\\
& &+W\sin\theta\cos(\varphi-\omega t)\nonumber\\
& &\dot\theta=P_{\theta}/(\xi+r_0)^2\mu\\ 
& &\dot P_{\theta}=\frac{P_{\varphi}^2\cos\theta}{\sin^3\theta (\xi+r_0)^2\mu}\\
& &+W(r_0+\xi)\cos\theta\cos(\varphi-\omega t)\nonumber \\ 
& &\dot\varphi=P_{\varphi}/\sin^2\theta (\xi+r_0)^2\mu\\
& &\dot P_{\varphi}=-W\sin\theta\sin(\varphi-\omega t) 
\end{eqnarray}
\end{subequations}
and
\begin{subequations}
\begin{eqnarray}
& &\dot\xi=P_{\xi}/\mu\\ \nonumber\\
& &\dot P_{\xi}=-\mu\omega_o^2\xi+3a^3D\xi^2-\frac{7}{3}a^4D\xi^3\\
& &-\frac{L^2}{2\mu r_0^2}\left[-\frac{2}{r_0}+
\frac{6\xi}{r_0^2}-\frac{12\xi^2}{r_0^3}+\frac{20\xi^3}{r_0^4}\right]\nonumber\\
& &\quad\quad\quad +W\sin\theta\cos(\varphi-\omega t)\\ \nonumber\\
& &\dot\theta=\frac{P_{\theta}}{r_0^2\mu}\left[1-\frac{2\xi}{r_0}+\frac{3\xi^2}{r_0^2}-\frac{4\xi^3}{r_0^3}+\frac{5\xi^4}{r_0^4}\right]\\ \nonumber\\
& &\dot P_{\theta}=\frac{P_{\varphi}^2\cos\theta}{\sin^3\theta r_0^2\mu}\left[1-\frac{2\xi}{r_0}+\frac{3\xi^2}{r_0^2}-\frac{4\xi^3}{r_0^3}+\frac{5\xi^4}{r_0^4}\right]\nonumber\\
& &\quad\quad\quad+W\xi\cos\theta\cos(\varphi-\omega t)\\ \nonumber\\
& &\dot\varphi=P_{\varphi}/\sin^2\theta r_0^2\mu\left[1-\frac{2\xi}{r_0}+\frac{3\xi^2}{r_0^2}-\frac{4\xi^3}{r_0^3}+\frac{5\xi^4}{r_0^4}\right]\\ \nonumber\\
& &\dot P_{\varphi}=-W\sin\theta\sin(\varphi-\omega t) 
\end{eqnarray}
\end{subequations}

\noindent
where $\omega_o=a \sqrt{2D/\mu}$. Equations (5) and (6) are solved through Runge-Kutta at fourth order method. We will focus our study with the molecules Berilyum Oxide ($BeO$),  Carbon Monoxide ($CO$) , and Cyanide ($CN$), where the parameters which characterize these molecules are shown on next table. The regular and chaotic behavior of the molecule is determined by the root square of its spectrum, $\sqrt{I(\nu)}$, that is, the Fourier transformation of its time evolution in the $\xi$ coordinate, although we also checked this behavior using the phase space picture ($\xi, P_{\xi}$), stroboscopic map of the space ($\xi, P_{\xi}$), and the Lyapunov's exponent ($\lambda$). The parameter $W^c$ shown on the table indicates the strength of the electric field for the system becomes chaotic ($W_e^c$ corresponds to the dynamical system (5), and $W_4^c$ corresponds to the dynamical system (6)). Figure 1 shows the squared root of the spectrum for the regular (only peaks) behavior of the molecules, and Figure 2 shows the   
chaotic (continuous component in the spectrum) behavior of the molecules.

\begin{table}[h!]
\centering
\begin{tabular}{|c |c| c| c|} 
 \hline\hline
                                                                                & $CO$  & $BeO$ & $CN$\\ [0.8ex] 
 \hline
  $D (eV)$                                                                      & $4.74$& $5.24$  & $9.49$ \\ 
 \hline
  $a$ $(1/\buildrel _\circ \over {\mathrm{A}})$                                 & $1.93$& $2.12$  & $2.32$   \\
  \hline
  $\mu$ $(u)$                                                                   & $6.85$& $5.76$ & $6.46$  \\ 
 \hline
  $\omega_o$ $ (\sqrt{eV/u}/\buildrel _\circ \over {\mathrm{A}})$               & $2.27$& $2.85$  & $3.97$   \\    
\hline
$r_0$ ($\buildrel _\circ \over {\mathrm{A}}$)                                   & $1.37$& $1.27$   & $1.14$   \\     
\hline
$W_e^c$($u \buildrel _\circ \over {\mathrm{A}}/s^2$)                            & $1.30$& $1.50$  & $3.50$  \\       
\hline
$W_4^c$($u \buildrel _\circ \over {\mathrm{A}}/s^2$)                            & $1.30$& $1.50$  & $3.50$   \\     
\hline\hline
\end{tabular}
\caption{Characteristics of the three diatomic molecules}
\label{table:1}
\end{table}

\noindent
For an amplitude of the electric field such that $W<W^c_{e,4}$, the behavior of the molecule is regular, and this is shown on Fig. 1  where the root square of the power spectrum, $\sqrt{I(\nu)}$ of the time depending variable $\xi(t)$ is presented. For and electric field magnitude such that $W\ge W^c_{e,4}$ the behavior of the molecule is chaotic, as it is shown on Fig. 2, where the plot shows a continuous component of the spectrum.  The critical value $W^c$ for onset the chaotic motions depends on the molecule,  and it does not depend on the approximation made, up to the third digit (which is not considered here). The critical value $W^c$ is denoted as $W^c_e$ for the exact case and   $W^c_4$ for the 4th-order approximation. These critical values are shown on the table. We need to point out that the onset of chaotic behavior of the molecules was also determined through Lyapunov's exponent \cite{31}, the stroboscopic map in the $\xi$-phase space \cite{32}, and detecting the sensibility of the system under initial conditions (not shown on this paper). 

\begin{figure}[h!]
{\centering
 \includegraphics[width=0.5\textwidth]{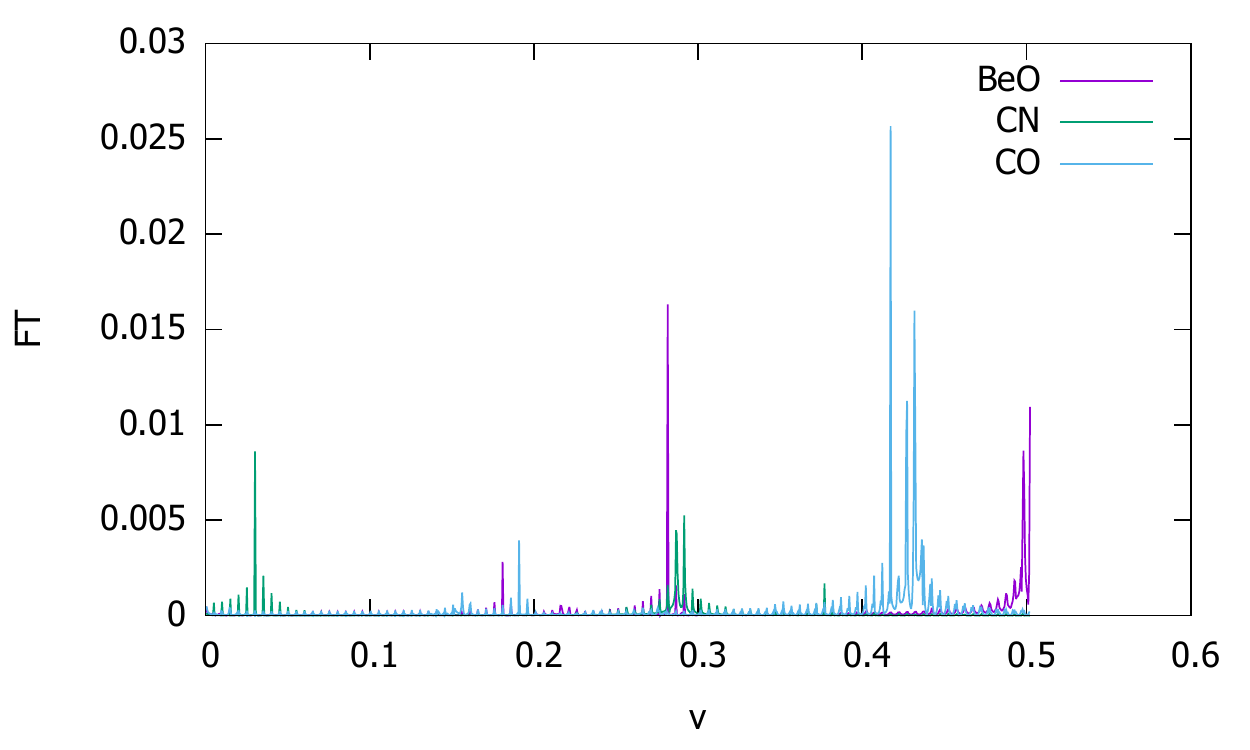}
 \caption{Regular behavior of diatomic molecules $BeO$,  $CO$, and  $CN$.}
 \label{RegularC1}}
 \end{figure} 

\begin{figure}[h!]
{\centering
 \includegraphics[width=0.5\textwidth]{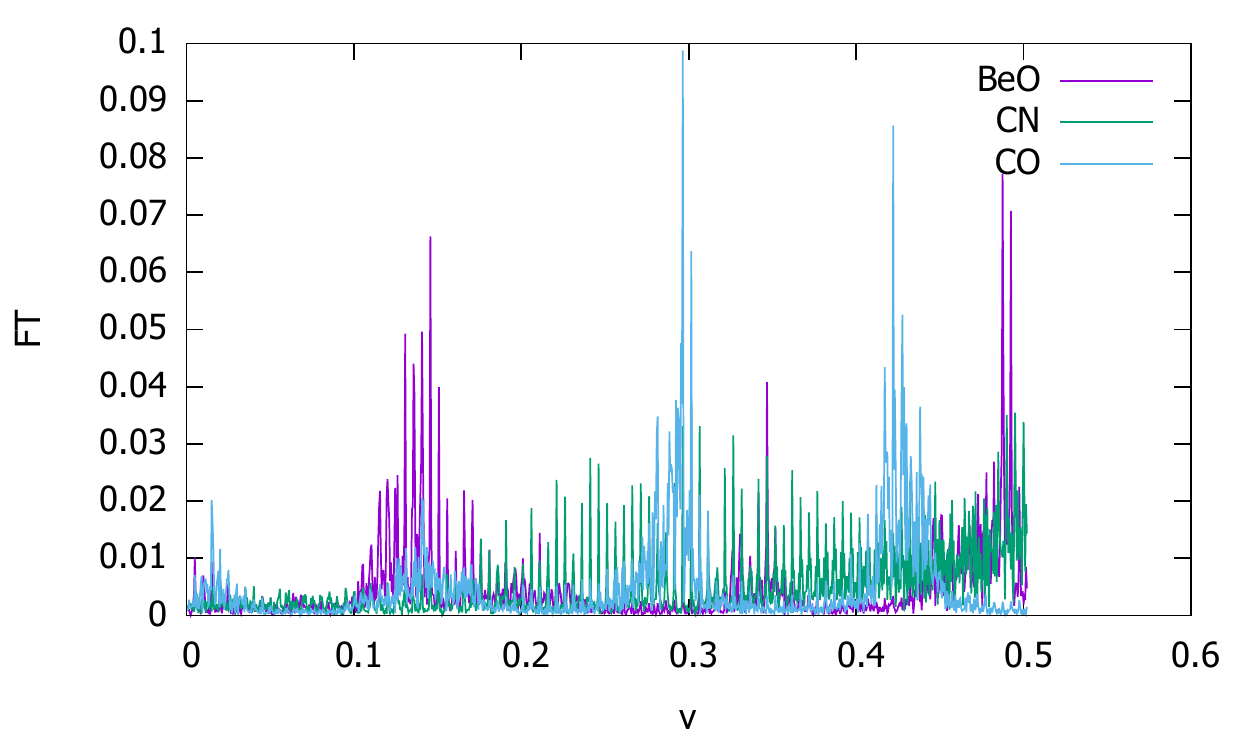}
 \caption{Chaotic behavior of diatomic molecules $BeO$, $CO$, and $CN$.}
 \label{ChaosC1}}
 \end{figure} 

\section{Quantum Analysis}
Due to the very good agreement of the dynamics with the fourth order approximation, we will use this approximation (\ref{ha4}) to study the quantum dynamics of the molecules and to solve the Sch\"odinger's equation
\begin{equation}\label{Sh4}
i\hbar\frac{\partial |\Psi\rangle}{\partial t}=\widehat{H}_4\;|\Psi\rangle,
\end{equation}
\noindent
where $\hbar$ is the (2$\pi$) Planck's constant , $\widehat{H}_4$ is the Hermitian linear operator associated to the classical Hamiltonian (\ref{ha4}), and $|\Psi\rangle$ is the wave function. This Hamiltonian can be written as a part which is independent on time plus a part that has, in addition, all time dependence,

\begin{equation}
\widehat{H}_4=\widehat{H}_{00}+\widehat{H}_{0}+\widehat{V}(t),
\end{equation} 
\noindent
where it follows that

\begin{subequations}

\begin{equation}
\widehat{H}_{00}=\frac{\widehat{P}_\xi^2}{2\mu}+\frac{1}{2}\mu\omega_o^2\xi^2+
\frac{\widehat{L}^2}{2\mu r_0^2}\;,
\end{equation}

\begin{eqnarray}
\widehat{H}_{0}=&\frac{\widehat{L}^2}{2\mu r_{0}^{2}}\left[ -\frac{2\xi}{r_{0}}+\frac{3\xi^{2}}{r_{0}^2}-\frac{4 \xi^{3}}{r_{0}^{3}}+\frac{5\xi^{4}}{r_{0}^4}\right] \\
&-a^{3}D\xi^{3}+\frac{7}{12}a^4D\xi^4\nonumber\;,
\end{eqnarray}
\noindent
and

\begin{equation}
\widehat{V}(t)= -W\xi\sin\theta\cos(\varphi-\omega t).
\end{equation}

\end{subequations}

Consider the expansion of the wave function of the wave function $|\Psi\rangle$ in terms of the basis $\{|nlm\rangle\}$,
\begin{equation}\label{wave}
|\Psi(t)\rangle=\sum_{n,l,m}C_{nlm}(t)e^{-i E_{nl}t/\hbar}|nlm\rangle,\quad\hbox{with}\quad \sum_{nlm}|C_{nlm}(t)|^2=1,
\end{equation}
\noindent
where $n,l\in{\cal Z}^+$ and $-l\le m\le l$, and the basis is such that

\begin{equation}
\left(\frac{\widehat{P}_{\xi}^2}{2\mu}+\frac{1}{2}\mu\omega_o^2\widehat{\xi^2}\right)|nlm\rangle=\hbar\omega_o(n+1/2)|nlm\rangle,
\end{equation}

\begin{equation}
\widehat{L}^2|nlm\rangle=\hbar^2l(l+1)|nlm\rangle,
\end{equation}
\noindent
and the energies $E_{nl}$ are given by

\begin{align}
E_{nl}=&\hbar\omega_{o}(n+1/2)+\frac{\hbar^{2}}{2 \mu r_{o}^{2}}\;l(l+1)+\frac{3\hbar^{3}}{4\mu^{2} r_{o}^{4}\omega_{o}}\;l(l+1)\;(2n+1)\nonumber\\
       &+\left[\frac{7}{12}a^{4}De+\frac{5\hbar^{2}}{2 \mu r_{o}^{6}}\;l(l+1)\right]\left[\frac{\hbar}{2\mu\omega_{o}}\right]^{2} 3(2n^{2}+2n+1)\;.
\end{align}

\noindent
After substituting (\ref{wave}) in (\ref{Sh4}) , doing some rearrangements, and using the orthogonality relation $\langle n'l'm'|nlm\rangle=\delta_{n'n}\delta_{l'l}\delta_{m'm}$, the equation for the 
coefficients are
\begin{eqnarray}\label{cdy}
&i\hbar\omega\;\dot{C}_{n'l'm'}(\tau)=\\
&\sum_{n'l'm'\neq nlm}{e^{i\left( E_{n'l'}-E_{nl}\right)\tau/\hbar\Omega}}\;C_{nlm}(\tau)\bra{n'l'm'}\widehat{H}_{0}+\widehat{V}(\tau)\ket{nlm}\nonumber\;,
\end{eqnarray}
where we have made the definition $\tau=\omega t$.

\noindent Once the matrix elements are calculated, the final dynamical system is given by \\ (see appendix \ref{A})
\vspace{0.1cm}
\begin{widetext}
\begin{eqnarray}\label{goliath}
&i\hbar\omega\;\dot{C}_{n'l'm'}(\tau)=A_{1}\;C_{n'+1\;l'\;m'}\;e^{i\left(\frac{E_{n'l'}-E_{n'+1\;l'}}{\hbar\omega}\right)\tau}+A_2\;C_{n'-1\;l'\;m'}\;e^{i\left(\frac{E_{n'l'}-E_{n'-1\;l'}}{\hbar\omega}\right)\tau}\nonumber\\
&+A_{3}\;C_{n'+2\;l'\;m'}\;e^{i\left(\frac{E_{n'l'}-E_{n'+2\;l'}}{\hbar\omega}\right)\tau}+A_{4}\;C_{n'-2\;l'\;m'}\;e^{i\left(\frac{E_{n'l'}-E_{n'-2\;l'}}{\hbar\omega}\right)\tau}\nonumber\\
&+A_{5}\;C_{n'+3\;l'\;m'}\;e^{i\left(\frac{E_{n'l'}-E_{n'+3\;l'}}{\hbar\omega}\right)\tau}+A_{6}\;C_{n'-3\;l'\;m'}\;e^{i\left(\frac{E_{n'l'}-E_{n'-3\;l'}}{\hbar\omega}\right)\tau}\nonumber\\
&+A_{7}\;C_{n'+4\;l'\;m'}\;e^{i\left(\frac{E_{n'l'}-E_{n'+4\;l'}}{\hbar\omega}\right)\tau}+A_{8}\;C_{n'-4\;l'\;m'}\;e^{i\left(\frac{E_{n'l'}-E_{n'-4\;l'}}{\hbar\omega}\right)\tau}\nonumber\\
&+B_{1}\;C_{n'+1\;l'+1\;m'-1}\;e^{i\left(\frac{E_{n'l'}-E_{n'+1\;l'+1}}{\hbar\omega}-1\right)\tau}-B_{2}\;C_{n'+1\;l'-1\;m'-1}\;e^{i\left(\frac{E_{n'l'}-E_{n'+1\;l'-1}}{\hbar\omega}-1\right)\tau}\nonumber\\
&+B_{3}\;C_{n'+1\;l'-1\;m'+1}\;e^{i\left(\frac{E_{n'l'}-E_{n'+1\;l'-1}}{\hbar\omega}+1\right)\tau}-B_{4}\;C_{n'+1\;l'+1\;m'+1}\;e^{i\left(\frac{E_{n'l'}-E_{n'+1\;l'+1}}{\hbar\omega}+1\right)\tau}\nonumber\\
&+B_{5}\;C_{n'-1\;l'+1\;m'-1}\;e^{i\left(\frac{E_{n'l'}-E_{n'-1\;l'+1}}{\hbar\omega}-1\right)\tau}-B_{6}\;C_{n'-1\;l'-1\;m'-1}\;e^{i\left(\frac{E_{n'l'}-E_{n'-1\;l'-1}}{\hbar\omega}-1\right)\tau}\nonumber\\
&+B_{7}\;C_{n'-1\;l'-1\;m'+1}\;e^{i\left(\frac{E_{n'l'}-E_{n'-1\;l'-1}}{\hbar\omega}+1\right)\tau}-B_{8}\;C_{n'-1\;l'+1\;m'+1}\;e^{i\left(\frac{E_{n'l'}-E_{n'-1\;l'+1}}{\hbar\omega}+1\right)\tau}
\end{eqnarray}
\end{widetext}
\noindent The meaning of $|C_{nlm}(\tau)|^2$ is, of course, the probability of having the molecule in the sate $|nlm\rangle$ at the normalized time "$\tau$." The complex dynamical system represented  
by the equations (\ref{goliath}) is solved using Runge-Kutta method at fourth order.  Once the solution is obtained, the Wigner function (see appendix B) is calculated at the end of the considered evolution ($\tau=20$),
\begin{equation*}
W(\xi, p_{_\xi}, \tau)=\frac{1}{\pi\hbar}\int_{\Re}\Psi^*(\xi+s,\tau)\Psi(\xi-s,\tau)e^{-i2p_{_\xi} s/\hbar}ds,
\end{equation*}
or
\begin{eqnarray}
W(\xi,p_{_\xi},\tau)&=&\frac{1}{\hbar \pi}\sum_{nn'lm}C^*_{nlm}(\tau)C_{n'lm}(\tau)\sqrt{\frac{2^n n!}{2^{n'} n'!}}(-1)^{n'}\;\nonumber\\
& &\times L_{n}^{n'-n}\left[2 \left(\frac{\mu \omega_{o}}{\hbar^2}\xi^2+\frac{p_{_\xi}^2}{\hbar^2}\right)\right]\;\nonumber\\
& &\times \left(-\sqrt{\frac{\mu \omega_{o}}{\hbar}}\xi +\frac{ip_{_\xi}}{\hbar}\right)^{n'-n}\\
& &\times\sqrt{\frac{\mu \omega_{o}}{\hbar}}\;e^{-\left(\frac{\mu \omega_{o}}{\hbar}\xi^2+\frac{p_{_\xi}^2}{\hbar}\right)}\nonumber\;,
\end{eqnarray}
\noindent
and the Boltzmann-Shannon (BS) entropy is also calculated,
\begin{equation}
S(\tau)=-\sum_{nlm}|C_{nlm}(\tau)|^2\ln|C_{nlm}(\tau)|^2,
\end{equation}
with its average value,
\begin{equation}\label{avS}
\langle S(\tau)\rangle=\frac{1}{T}\int_0^TS(\tau)d\tau,
\end{equation}

\vspace{0.1cm}
\noindent where $T$ is the total normalized time evolution of the quantum system.

\section{Numerical Results of the Quantum Case}
Using the data shown on above table for the molecules $BeO$, $CO$ and $CN$, we  solve the dynamical system (\ref{cdy}) for $l=2$, and $-2\le m\le 2$ and n=4, having a total number of forty five quantum states.
The initial conditions were chosen as $C_{000}(0)=1$, $C_{nlm}(0)=0$ for $n\not=0$, $l\not=0$ , and $m\not=0$ (the system is initially in the ground state). We select the non resonant case $\omega=\omega_o$ with $\varphi=0$ as the characteristics of the electric field, and we chose the normalized time $\tau=20$ (end of the evolution) to calculate the Wigner function. For an electric strength such that $W< W_4^c$, the Wigner function is shown on Figures 3, which correspond to the regular classical dynamics. As one see,  the system tends to be in the origin of the phase space (maximum of the Wigner function).
\begin{figure*}
{\centering
 \includegraphics[width=1.0\textwidth]{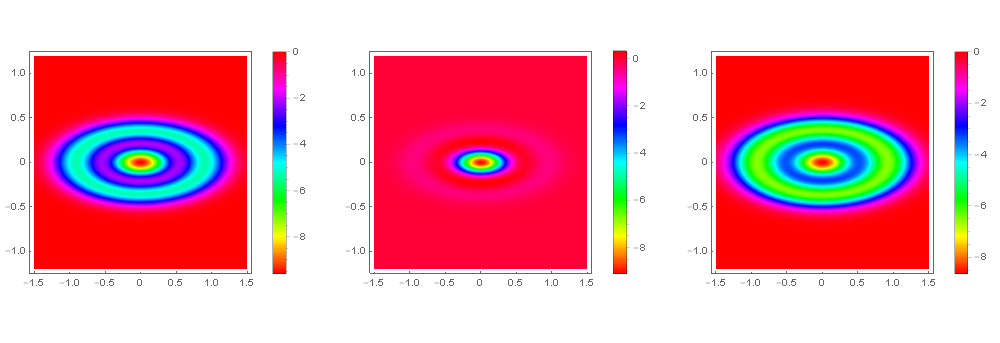}
 \caption{Wigner function of the molecules $CO$, $BeO$, $CN$ for $W=0.1$ (from left to right)}
 \label{RegularC1}}
 \end{figure*} 

\noindent
For an electric field strength $W\ge W_4^c$, the Wigner function and the BS entropy are shown on Figures 4 and 5 below.  As one can see, the system try to move away from the center of the phase space, forming a very local minimum of the Wigner function. 
\begin{figure*}
{\centering
 \includegraphics[width=1.0\textwidth]{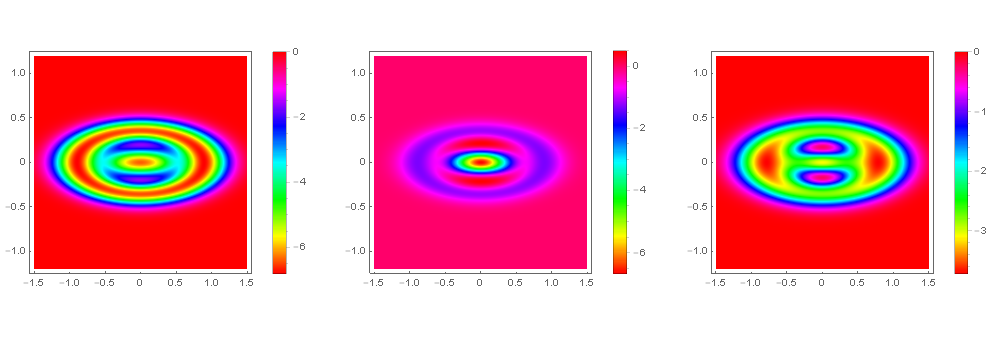}
 \caption{Wigner function for CO ($W^c=1.3$), BeO ($W^c=1.5$), and CN ($W^c=3.5$).}
 \label{RegularC1}}
 \end{figure*} 
\begin{figure*}
{\centering
 \includegraphics[width=1.0\textwidth]{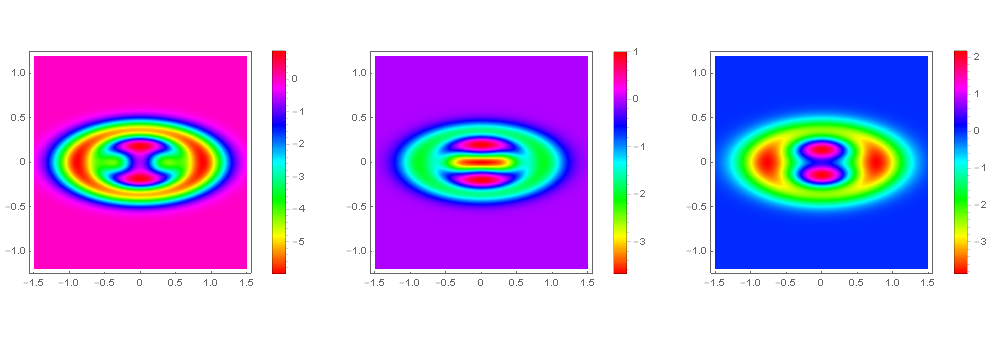}
 \caption{Wigner function for CO ($W=3.0$), BeO ($W=2.5$), and CN ($W=5.0$).}
 \label{RegularC1}}
 \end{figure*} 

Figure 6 shows the BS entropy for the $BeO$ molecule for the regular classical behavior ($W=0.1$), for its critical behavior ($W=1.5$), and for its upper critical behavior ($W=2.5$).  Classical regular behavior corresponds to almost constant BS entropy (after a fast increasing) in the quantum case, meanwhile classical chaotic behavior corresponds to almost a linear increasing in BS entropy (after the same fast increasing), that is, more quantum information is lost or the quantum system become more disordered (more states increases their role in the quantum dynamics).

Figure 7 shows the average value of the BS entropy as a function of the electric field strength $W$, with and normalized time $T=20$. The first jump occurs just at $W=W_4^c$ found in the classical case when chaotic classical dynamics starts. Then, a quantum signature of this classical phenomenon correspond to a sudden increase of the BS entropy (sudden lost of quantum information). However, there are another jumps for $W>W_4^{c}$, indicating that another sudden lost of information has occurred at this value. In turns, this also indicates that at the classical case there must be a different chaotic behavior from the previous already seen. 
To check this implication, we calculate the squared root of the power spectrum, $\sqrt{I(\nu)}$, for the $BeO$ molecule and for the cases: $1.5 \le W< 2.5$ (first chaotic behavior), $2.5 \le W< 5.5$ (second chaotic behavior), and  for $W\ge 5.5$ (third chaotic behavior), and  the result are shown on Figure 8. As one can see, effectively, the there are a second and third chaotic classical behaviors  which have a wider continuous component in the spectrum (we point our that the same feature was observed with the $CN$ and $CO$ molecules, not shown on this paper).  

\begin{figure}[h!]
{\centering
 \includegraphics[width=0.5\textwidth]{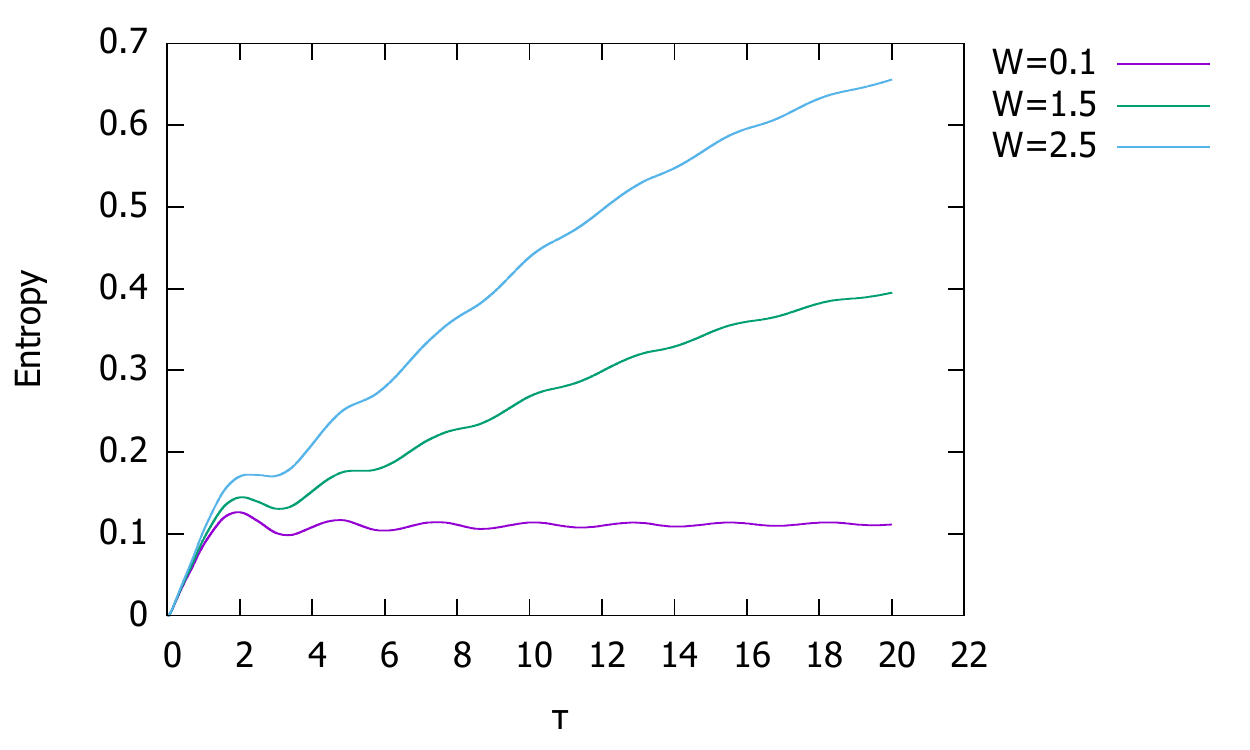}
 \caption{BS entropy for the $BeO$ molecule. }
 \label{ChaosC1}}
 \end{figure} 

\begin{figure}[h!]
{\centering
 \includegraphics[width=0.5\textwidth]{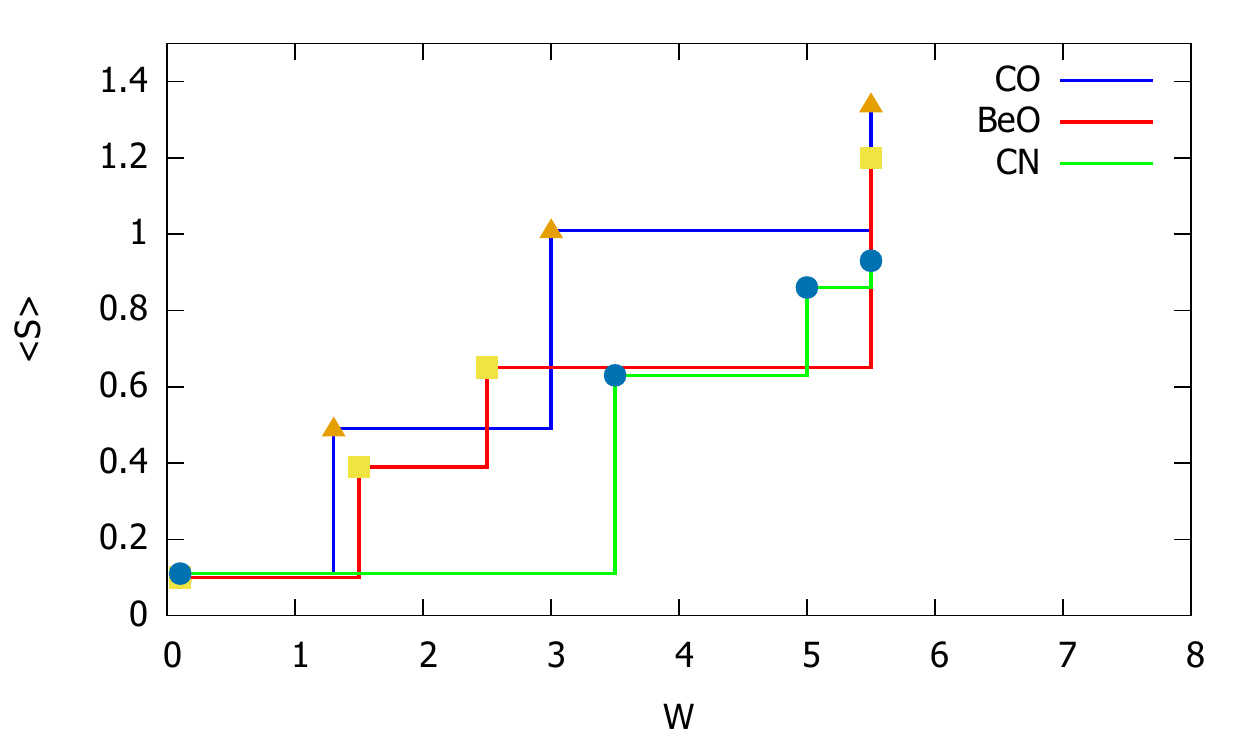}
 \caption{Average time of BS entropy versus electric field amplitude. }
 \label{ChaosC2}}
 \end{figure} 

\noindent
Then this result indicates that one can use average BS entropy to determine the magnitude of the parameter $W$ at which the classical chaos may appears. In addition and due to other jumps on the average BS entropy, it seems that it can help to classify different type of classical chaos of the system. To see this, we compared the chaotic behavior of the $BeO$ molecule at $W^c=1.5, 2.5$, and $5.5$. For these two former  values, the classical system is more sensitive to initial conditions (not shown here), a positive Lyapunov's exponent appears  earlier (not shown here), the Wigner function is totally different (see Figures 4 and 5), and the spectrum has more continuous component (see next Figure) compared  with the first chaotic value $W=1.5$.
\begin{figure}[h!]
{\centering
 \includegraphics[width=0.5\textwidth]{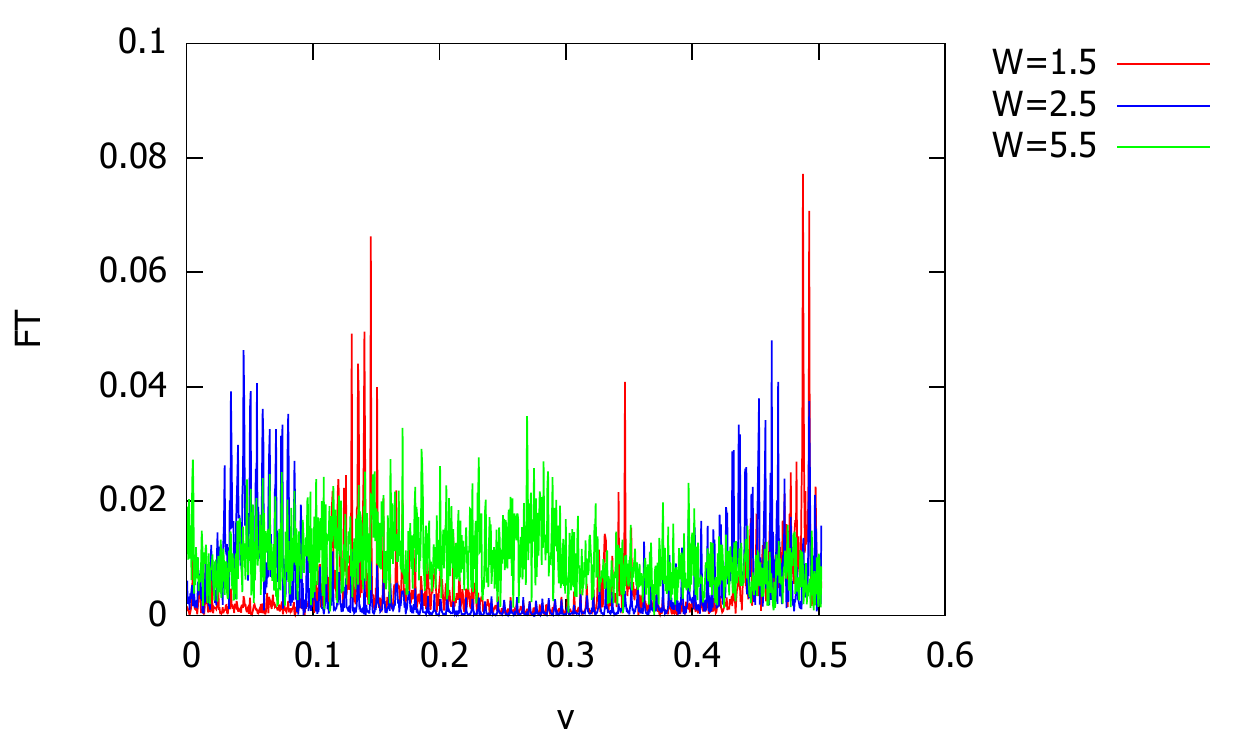}
 \caption{Chaotic behavior comparison for the $BeO$ molecule.}
 \label{ChaosC2}}
 \end{figure} 

\section{Conclusions}
We have studied the classical chaotic behavior of the diatomic molecules $BeO$, $CO$ and $CN$ under a polarized time depending electric field, where the critical field strength $W^c$ was found with the exact Morse's potential ($W_e^c$), and with its approximation at fourth order ($W_4^c$). Both values are close enough to consider that the approximation at fourth order is very good to study the quantum behavior of the molecules with it. We made the quantization of the system within this approximation and used the Wigner function and BS entropy to see their quantum signature when the classical system becomes chaotic. We observed that the maximum of the Wigner function  move away  from the center of the phase space when the classical system becomes chaotic. On the other hand, the BS entropy has a linear increasing with respect to time when the classical system becomes chaotic, and it is almost constant when the classical system has a regular behavior. In addition, the average  BS entropy suffers a jump just at the $W_4^c$ value, when the classical system becomes chaotic. This indicates a sudden lost of quantum information, and as  the electric field strength becomes higher than $W_4^{c}$,  another jumps of the average BS entropy are found, indicating a further lost in quantum information, and the existence of qualitatively different classical chaotic behaviors  in the classical case. This was stated here using the spectrum of the system, where a qualitatively different spectra were found with wider continuous components. This study suggests that the average BS entropy in the quantum system could be used to determine the critical value for the classical system to becomes chaotic, and also it suggests that the average BS entropy could be used  to determine different classical chaotic behaviors.

\vspace{-.4cm}

\begin{acknowledgments}
\noindent We wish to acknowledge to CONACyT $572988$ for \\
support this proyect. 
\end{acknowledgments}

\appendix
\section{\textsc{Coefficients appearing on dynamical system}}\label{A}
\noindent The A's and B's constants named before have the\\
following values according to the matrix 
elements:

\begin{align*}
A_{1}&=-\frac{\hbar^2}{\mu r_{o}^3}l'(l'+1)\sqrt{\frac{\hbar(n'+1)}{2\mu\omega_{o}}}\\
&-\left[\frac{2\hbar^2}{\mu r_{o}^5}l'(l'+1)+a^3D\right]\left[\frac{\hbar}{2 \mu \omega_{o}}\right]^{3/2}\;3(n'+1)\sqrt{n'+1}\\
A_{2}&=-\frac{\hbar^2}{\mu r_{o}^3}l'(l'+1)\sqrt{\frac{\hbar\; n'}{2\mu\omega_{o}}}\\
&-\left[\frac{2\hbar^2}{\mu r_{o}^5 }l'(l'+1)+a^3D\right]\left[\frac{\hbar}{2 \mu \omega_{o}}\right]^{3/2}\;3\;n'\sqrt{n'}\\
A_{3}&=\frac{3}{4}\frac{\hbar^3}{\omega_{o}\mu^{2}r_{0}^4}l'(l'+1)\;\sqrt{(n'+2)(n'+1)}+\\
&+\left[\frac{5\hbar^2}{2\mu r_{0}^6}l'(l'+1)+\frac{7}{12}a^4D\right]\left[\frac{\hbar^2}{2\mu \omega_{o}}\right]^2\\
&\times\sqrt{(n'+2)(n'+1)}(4n'+6)\\
A_{4}&=\frac{3}{4}\frac{\hbar^3}{\omega_{o}\mu^{2}r_{0}^4}l'(l'+1)\;\sqrt{n'\;(n'-1)}\\
&+\left[\frac{5\hbar^2}{2\mu r_{0}^6}l'(l'+1)+\frac{7}{12}a^4D\right]\left[\frac{\hbar^2}{2\mu \omega_{o}}\right]^2\\
&\times\sqrt{n'\;(n'-1)}(4n'-2)\\
A_{5}&=-\left[\frac{2\hbar^2}{\mu r_{o}^5}l'(l'+1)+a^3D\right]\left[\frac{\hbar}{2 \mu \omega_{o}}\right]^{3/2}\\
&\times\sqrt{(n'+3)(n'+2)(n'+1)}\\
A_{6}&=-\left[\frac{2\hbar^2}{\mu r_{o}^5}l'(l'+1)+a^3D\right]\left[\frac{\hbar}{2 \mu \omega_{o}}\right]^{3/2}\\
&\times\sqrt{n'\;(n'-1)(n'-2)}\\
A_{7}&=\left[\frac{5\hbar^2}{2\mu r_{0}^6}l'(l'+1)+\frac{7}{12}a^4D\right]\left[\frac{\hbar^2}{2\mu \omega_{o}}\right]^2\\
&\times\sqrt{(n'+4)(n'+3)(n'+2)(n'+1)}\\
A_{8}&=\left[\frac{5\hbar^2}{2\mu r_{0}^6}l'(l'+1)+\frac{7}{12}a^4D\right]\left[\frac{\hbar^2}{2\mu \omega_{o}}\right]^2\\
&\times\sqrt{n'(n'-1)(n'-2)(n'-3)}
\end{align*}

\begin{align*}
B_{1}&=\frac{W}{2}\sqrt{\frac{\hbar}{2\mu\omega_{o}}}\sqrt{n'+1}\sqrt{\frac{(l'-m'+2)(l'-m'+1)}{(2l'+3)(2l'+1)}}\\
B_{2}&=\frac{W}{2}\sqrt{\frac{\hbar}{2\mu\omega_{o}}}\sqrt{n'+1}\sqrt{\frac{(l'+m'-1)(l'+m')}{(2l'-1)(2l'+1)}}\\
B_{3}&=\frac{W}{2}\sqrt{\frac{\hbar}{2\mu\omega_{o}}}\sqrt{n'+1}\sqrt{\frac{(l'-m'-1)(l'-m')}{(2l'-1)(2l'+1)}}\\
B_{4}&=\frac{W}{2}\sqrt{\frac{\hbar}{2\mu\omega_{o}}}\sqrt{n'+1}\sqrt{\frac{(l'+m'+1)(l'+m'+2)}{(2l'+3)(2l'+1)}}\\
B_{5}&=\frac{W}{2}\sqrt{\frac{\hbar}{2\mu\omega_{o}}}\sqrt{n'}\sqrt{\frac{(l'-m'+2)(l'-m'+1)}{(2l'+3)(2l'+1)}}\\
B_{6}&=\frac{W}{2}\sqrt{\frac{\hbar}{2\mu\omega_{o}}}\sqrt{n'}\sqrt{\frac{(l'+m'-1)(l'+m')}{(2l'-1)(2l'+1)}}\\
B_{7}&=\frac{W}{2}\sqrt{\frac{\hbar}{2\mu\omega_{o}}}\sqrt{n'}\sqrt{\frac{(l'-m'-1)(l'-m')}{(2l'-1)(2l'+1)}}\\
B_{8}&=\frac{W}{2}\sqrt{\frac{\hbar}{2\mu\omega_{o}}}\sqrt{n'}\sqrt{\frac{(l'+m'+1)(l'+m'+2)}{(2l'+3)(2l'+1)}}
\end{align*}
%

\section{Wigner distribution function}\label{B}
The Wigner distribution is given by
\begin{align*}
&W(X,P,\tau)=\\
           &\frac{1}{\hbar \pi}\int_{\Omega\times \mathbb{R}}{\Psi^*(X+Y,\tau)\Psi(X-Y,\tau)e^{2iPY/\hbar}dY\;d\Omega}\\
           &=\frac{1}{\hbar \pi}\sum_{nlm\neq n'l'm'}C^*_{nlm}(\tau)C_{n'l'm'}(\tau)\int_{\Omega}{Y^*_{lm}Y_{l'm'}d\Omega}\\
					 &\times\int_{\mathbb{R}}{\psi^*_{n}(X+Y)\psi_{n'}(X-Y)e^{2iPY/\hbar}dY}\\
           &=\frac{1}{\hbar \pi}\sum_{nn'lm}C^*_{nlm}(\tau)C_{n'lm}(\tau)\\
					 &\times\int_{\mathbb{R}}{\psi^*_{n}(X+Y)\psi_{n'}(X-Y)e^{2iPY/\hbar}dY}\;.
\end{align*}

\noindent where $\Psi=C_{nlm}(\tau)\;Y_{nlm}\;\psi_{n}$ with $Y_{nlm}$ as spherical harmonics and $\psi_{n}$ the wave function of harmonic oscillator.

\noindent  The wave function for harmonic oscillator is
\begin{equation*}
\psi_n=\frac{1}{\sqrt{2^n n!}}\;\left({\frac{\mu \omega_{o}}{\hbar \pi}}\right)^{1/4}e^{-\frac{\mu \omega_{o}}{2 \hbar}X^2} H_{n}\left(\sqrt{\frac{\mu \omega_{o}}{\hbar}}X\right)\;,
\end{equation*}
\noindent where $H_{n}$ are the Hermit polynomials. Let $\tilde X$  and $a_n$ be defined as

\begin{align*}
\tilde{X}&=\sqrt{\frac{\mu \omega_{o}}{\hbar}}X\\
    a_{n}&=\frac{1}{\sqrt{2^n n!}}\;\left({\frac{\mu \omega_{o}}{\hbar \pi}}\right)^{1/4}.
\end{align*}
\noindent Then, the wave function can be written as

\begin{equation*}
\psi_{n}(X)=a_{n}\; e^{-\frac{\tilde{X}^2}{2}}H_{n}(\tilde{X})\;.
\end{equation*}

\noindent In order to know the complete form of Wigner distribution we have to calculate 
\begin{equation*}
\int_{\mathbb{R}}{\psi^*_{n}(X+Y)\psi_{n'}(X-Y)e^{2iPY/\hbar}dY}\;,
\end{equation*}
with the above harmonic oscillator's wave function.

\begin{align*}
&\int_{\mathbb{R}}{\psi^*_{n}(X+Y)\psi_{n'}(X-Y)e^{2iPY/\hbar}dY}\\
&=e^{-\tilde{X}^2}\int_{-\infty}^{\infty}{e^{-Y^2}H_n(\tilde{X}+Y)H_{n'}(\tilde{X}-Y)e^{2iPY/\hbar}dY}\\
&= e^{-\tilde{X}^2}\int_{-\infty}^{\infty}{e^{-[(Y-iP/\hbar)^2+P^2/\hbar^2]}H_n(\tilde{X}+Y)H_{n'}(\tilde{X}-Y)dY}\\
&= e^{-(\tilde{X}^2+P^2/\hbar^2)}\\
&\times\int_{-\infty}^{\infty}{e^{-(Y-iP/\hbar)^2}H_n(\tilde{X}+Y)H_{n'}(\tilde{X}-Y)dY}\;,
\end{align*}
\noindent
Let $Z$ be defined as $Z=Y-iP/\hbar$. So, the last integral can be written as

\begin{equation*}
\int_{-\infty}^{\infty}{e^{-Z^2}H_n(\tilde{X}+Z+iP/\hbar)H_{n'}(\tilde{X}-Z-iP/\hbar)dZ}.
\end{equation*}
Using the relation $H_{n}(-x)=(-1)^n\;H_{n}(x)$, we have

\begin{equation*}
\int_{-\infty}^{\infty}{e^{-Z^2}H_n(Z+\tilde{X}+iP/\hbar)H_{n'}(Z-\tilde{X}+iP/\hbar)dZ}
\end{equation*}
\noindent
With $\alpha_1$ and $\alpha_2$ being defined as $\alpha_1=\tilde{X}+iP/\hbar$\; and \;$\alpha_2=-\tilde{X}+iP/\hbar$\;, it follows that

\begin{align*}
&\int_{-\infty}^{\infty}{e^{-Z^2}H_n(Z+\alpha_1)H_{n'}(Z+\alpha_2)dZ}=\\
&(-1)^{n'} 2^{n} \pi n!\;\alpha_{2}^{n'-n}\;L_{n}^{n'-n}(-2\alpha_1 \alpha_2)
\end{align*}
\noindent which is given in terms of our original variables as

\begin{align*}
&\int_{\mathbb{R}}{\psi^*_{n}(X+Y)\psi_{n'}(X-Y)e^{2iPY/\hbar}dY}=\\
&\sqrt{\frac{2^n n!}{2^{n'} n'!}}(-1)^{n'} L_{n}^{n'-n}\left[2 \left(\frac{\mu \omega_{o}}{\hbar^2}X^2+\frac{P^2}{\hbar^2}\right)\right]\;\\
&\times \left(-\sqrt{\frac{\mu \omega_{o}}{\hbar}}X +\frac{iP}{\hbar}\right)^{n'-n} \sqrt{\frac{\mu \omega_{o}}{\hbar}}\;\frac{1}{\hbar \pi}e^{-\left(\frac{\mu \omega_{o}}{\hbar}X^2+\frac{P^2}{\hbar}\right)}
\end{align*}

\vspace{0.1cm}

\noindent Finally, the Wigner is written in terms of the wave coefficients as
\begin{widetext}
\begin{align*}
&W(X,P,\tau)=\frac{1}{\hbar \pi}\sum_{nn'lm}C^*_{nlm}(\tau)C_{n'lm}(\tau)\sqrt{\frac{2^n n!}{2^{n'} n'!}}(-1)^{n'} L_{n}^{n'-n}\left[2 \left(\frac{\mu \omega_{o}}{\hbar^2}X^2+\frac{P^2}{\hbar^2}\right)\right]\\
&\times\left(-\sqrt{\frac{\mu \omega_{o}}{\hbar}}X +\frac{iP}{\hbar}\right)^{n'-n} \sqrt{\frac{\mu \omega_{o}}{\hbar}}\;e^{-\left(\frac{\mu \omega_{o}}{\hbar}X^2+\frac{P^2}{\hbar}\right)}\;.
\end{align*}
\end{widetext}


\begin{thebibliography}{99}
 

	\bibitem{1}
  A.J. Lichtenberg and M.A. Liberman,
  \emph{Regular and Stochastic Motion},
  Springer-Verlag, Berlin,
  (1983).
  
  \bibitem{2}
  G. Casati, B.V. Chirikov, D.L. Shepelyansky, and I. Guarnery,
  \emph{}
  Phys. Rep., {\bf 154}, 77,
  (1983).
  
  \bibitem{3}
  P. Labastie, M.C. Bordas, B. Tribollet, and M. Broyer,
  \emph{}
  Phys. Rev. Lett., {\bf 52}, 1681,
  (1984).
  
  \bibitem{4}
 J. Chevaleyre, C. Bordas, M. Broyer, and P. Labastie,
  \emph{}
 Phys. Rev. Lett., {\bf 57}, 3027,
  (1986).
  
  \bibitem{5}
  C. Bordas, P.F. Brevet, M. Broyer, J. Chevaleyre, P. Labastie, and J.P. Perrot,
  \emph{}
  Phys. Rev. Lett., {\bf 60}, 917,
  (1988).
  
  \bibitem{6}
  M. Lombardi, P. Labastie, M.C. Bordas, and M. Boyer,
  \emph{},
  J. Chem. Phys., {\bf 89}, 3479,
  (1988).
  
  \bibitem{7}
  M. Lombardi and T.H. Seligman,
  \emph{}
 Phys. Rev. A, {\bf 47}, 3571,
  (1993).
  
  \bibitem{8}
  J.J. Kay, S.L. Coy, V.S. Petrovi\'c, B.M. Wong, and R.W. Field,
  \emph{}
  J. Chem. Phys., {\bf 128}, 194301,
  (2008).
  
  \bibitem{9}
  D. Sugny, L. Bomble, T. Ribeyre, O. Dulieu, and M. Desouter-Lecomte,
  \emph{}
  Phys. Rev. A, {\bf 80}, 042325,
  (2009).
  
  \bibitem{10}
  A. Ruiz, J.P. Palao, and E.J. Heller,
  \emph{}
 Phys. Rev. E, {\bf 80}, 066606,
  (2009).
  
  
  \bibitem{11}
  B.V. Chirikov,
  \emph{}
  Phys. Rep. {\bf 52}, 263,
  (1979).
  
  \bibitem{12}
  \'E. V. Shuryak,
  \emph{}
  Sov. Phys. JEPT, {\bf 44}, 1070,
  (1976).
  
  \bibitem{13}
  R.P. Parson,
  \emph{}
  J. Chem. Phys. {\bf 88}, 3655,
  (1987).
  
  
  \bibitem{14}
  P.S. Dardi and K. Gray
  \emph{}
  J. Chem. Phys. {\bf 77}, 1345,
  (1982).
  
  \bibitem{15}
  G.P. Berman and A.R. Kolovsky,
  \emph{}
  Sov. Phys. JEPT {\bf 68}, 898,
  (1989).
  
   \bibitem{16}
  G.P. Berman and A.R. Kolovsky,
  \emph{}
  Sov. Phys. Usp. {\bf 35}, 303,
  (1992).

\bibitem{17}
  P. M. Morse,
  \emph{}
  Phys. Rev. {\bf 34}, 57
  (1929).
  
  
\bibitem{18}
  G. P. Berman, E. N. Bulgakov and D. D. Holm,
  \emph{}
  Phys. Rev. A {\bf 52}, 3074
  (1995).

\bibitem{19}
  B.V. Chirikov,
  Phys. Rep. {\bf 52}, 263
  (1979).

\bibitem{20}
  Ciann-Dong Yang and Hung-Jen Weng,
  Chaos and Fractral {\bf 38}, 16
  (2008).
  
  \bibitem{21}
  N.Moiseyev, H.J. Korsch, and B. Mirbach,
 Z. f\"ur Physik D Atoms, Molecules and Clusters, {\bf 29}, 125
  (1994).
  
  \bibitem{22}
  J. Gong and P. Brumer,
  J. Chem. Phys.  {\bf 115}, 3590
  (2001).
  
  \bibitem{23}
  S. Guruparan, B.R.D. Nayagam, S. Selvaraj, V. Ravichandran, and V. Chinnathambi,
  J. Adv. Chem. Scie.  {\bf 2}, 188
  (2016).
  
  \bibitem{24}
  C.A. Arango, W.W. Kennerly, and G.S. Ezra,
  J. Chem. Phys.  {\bf 122}, 184303
  (2005).
  
  \bibitem{25}
  M.C. Gutzwiller,
  {\it Chaos in Classical and Quantum Mechanics}, (1990), Springer-Verlag. 
 
  \bibitem{26}
  T.H. Seligman,
  J. Phys. A, {\bf 18}, (1985), 2751.

 \bibitem{27}
  O. Bohigas, M.J. Giannoni, and C. Schmit,
  Phys. Rev. Lett., {\bf 52}, (1984), 1.
  
  \bibitem{28}
   E. P. Wigner, Phys. Rev. {\bf 40}, (1932), 749.
  
  \bibitem{29}
  L. Boltzmann , Wiener Berichte. {\bf 53}, (1866), 195.
  
  \bibitem{30}
  C.E. Shannon, 
  The  Bell  System Technical Journal {\bf 27}, (1948), 379  and 623.
  
 \bibitem{31}
 P.G. Drazin,
 {\it Nonlinear Systems}, (1992), Cambridge University Press.
 
 \bibitem{32}
 R.C. Hilborn,
 {\it Chaos and Nonlinear Dynamics}, (1994), Oxford University Press.

\end{thebibliography}
\end{document}